\documentclass[prd,twocolumn,aps,superscriptaddress,a4paper]{revtex4-1}

\usepackage{color}
\usepackage{dcolumn}
\usepackage{amsmath}
\usepackage{amssymb}
\usepackage{mathtools}
\usepackage{graphicx}
\usepackage{latexsym}
\usepackage{amsfonts}
\usepackage{amssymb}
\usepackage{comment}
\usepackage[colorlinks=true,pdfstartview=FitV,linkcolor=black,citecolor=black,urlcolor=blue]{hyperref}
\DeclareGraphicsExtensions{.pdf,.gif,.jpg}

\newcommand{\be}{\begin{equation}}  
\newcommand{\ee}{\end{equation}}
\newcommand{\beq}{\begin{eqnarray}}  
\newcommand{\eeq}{\end{eqnarray}}  

\def\hc{\text{h.c.}}

\def\w{{\omega}}

\def\a{\alpha}  
  
\def\g{\gamma}

\def\eps{\epsilon}

\begin{document}

%%%%%%%%%%%%%%%%%%%%%%%%%%%%%%%%%%%%%%%%%%%%%%%%%%%%%%%%%%%%%%%%%%%%%%%%%%%%%%%%
%%%%%%%%%%%%%%%%%%%%%%%%%%%%%%%%%%%%%%%%%%%%%%%%%%%%%%%%%%%%%%%%%%%%%%%%%%%%%%%%

% Title stuff
\title{Time-resolved impurity-invisibility in graphene nanoribbons}

\author{Riku Tuovinen}
\email{riku.tuovinen@mpsd.mpg.de}
\affiliation{Max Planck Institute for the Structure and Dynamics of Matter, 22761 Hamburg, Germany}

\author{Michael A. Sentef}
\affiliation{Max Planck Institute for the Structure and Dynamics of Matter, 22761 Hamburg, Germany}

\author{Claudia Gomes da Rocha}
\affiliation{Department of Physics and Astronomy, University of Calgary, 2500 University Drive NW, Calgary, Alberta T2N 1N4, Canada}

%\affiliation{School of Physics and Centre for Research on Adaptive Nanostructures and Nanodevices (CRANN), Trinity College Dublin, Dublin 2, Ireland}

\author{Mauro S. Ferreira}
\affiliation{School of Physics, Trinity College Dublin, Dublin 2, Ireland}
\affiliation{Centre for Research on Adaptive Nanostructures and Nanodevices (CRANN), Trinity College Dublin, Dublin 2, Ireland}

\date{\today}  

% Abstract
\begin{abstract}

We investigate time-resolved charge transport through graphene nanoribbons supplemented with adsorbed impurity atoms. Depending on the location of the impurities with respect to the hexagonal carbon lattice, the transport properties of the system may become invisible to the impurity due to the symmetry properties of the binding mechanism. This motivates a chemical sensing device since dopants affecting the underlying sublattice symmetry of the pristine graphene nanoribbon introduce scattering. Using the time-dependent Landauer--B{\"u}ttiker formalism, we extend the stationary current-voltage picture to the transient regime, where we observe how the impurity invisibility takes place at sub-picosecond time scales further motivating ultrafast sensor technology. We further characterize time-dependent local charge and current profiles within the nanoribbons, and we identify rearrangements of the current pathways through the nanoribbons due to the impurities. We finally study the behavior of the transients with ac driving which provides another way of identifying the lattice-symmetry breaking caused by the impurities.

\end{abstract}

  
\maketitle

%%%%%%%%%%%%%%%%%%%%%%%%%%%%%%%%%%%%%%%%%%%%%%%%%%%%%%%%%%%%%%%%%%%%%%%%%%%%%%%%
%%%%%%%%%%%%%%%%%%%%%%%%%%%%%%%%%%%%%%%%%%%%%%%%%%%%%%%%%%%%%%%%%%%%%%%%%%%%%%%%

\section{Introduction}\label{sec:intro}
Being under considerable research focus for the past two decades graphene~\cite{Novoselov2005} and carbon nanotubes~\cite{Iijima1991} have been found to be extremely sensitive to external perturbations. For this reason, these nanomaterials have been proposed as ideal candidates for sensor technology~\cite{Gruner2006,Rodrigo2015,Goldsmith2019}. Based on these observations, carbon-based sensor devices have already been developed at the single-molecule resolution~\cite{Besteman2003,So2005,Choi2013,Shivananju2016}. Carbon-based transducers have been embedded in circuitries involving graphene nanopore platforms~\cite{Merchant2010,Avdoshenko2013} and field-effect transistors~\cite{Choi2012}, and these have been successfully applied to, e.g., disentangle biomolecules' rapid dynamics~\cite{Schneider2010,Garaj2010,Traversi2013}. This novel biosensor technology provides real-time information about the underlying physical and chemical mechanisms. These findings motivate ultrafast sensing devices as present transport measurements are able to resolve temporal information at sub-picosecond time scales~\cite{Prechtel2012,Hunter2015,Rashidi2016,Karnetzky2018,McIver2018}.

\begin{figure}[t]
\centering
\includegraphics[width=0.45\textwidth]{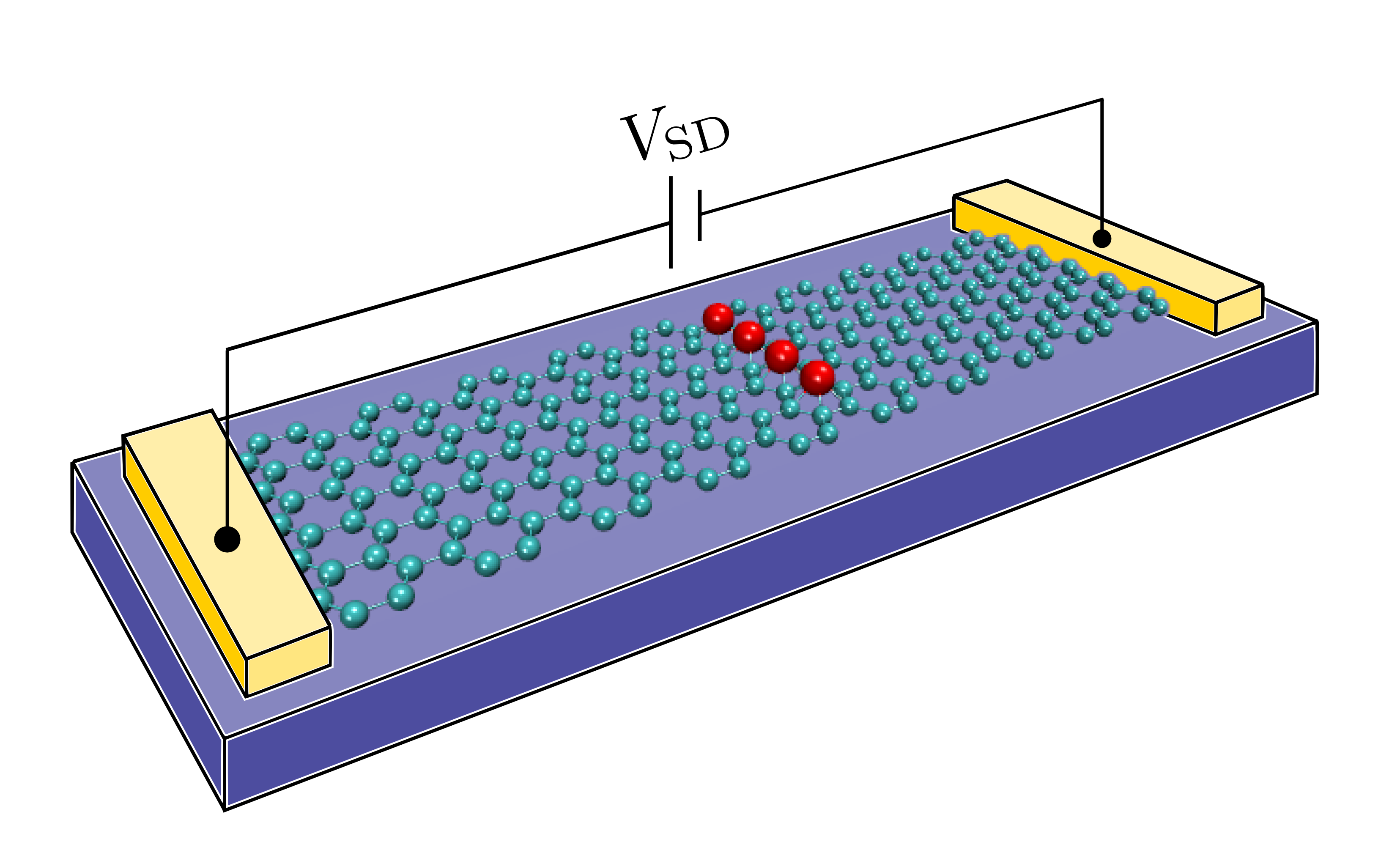}
\caption{Transport setup of an $N=11$ armchair graphene nanoribbon with adsorbed impurity atoms (red spheres) and $N$ indicating the number of carbon-dimers across the ribbon width. Contacts to the metallic leads are from the terminal sites of the nanoribbon. The leads are further connected to a source-drain voltage ($V_{SD}$).}
\label{fig:setup}
\end{figure}

These few examples include devices which are interacting with their persistently evolving environment for which a schematic in a quantum transport setup is shown in Fig.~\ref{fig:setup}. The schematics depicts a quantum transport channel made of a graphene nanoribbon (GNR) of armchair configuration with a width characterized by the number of carbon-dimers arranged transversely ($N$). The ribbon is subjected to a source-drain voltage ($V_{\text{SD}}$) that can vary on time and the whole device serves as a host for detecting the presence of impurities. The main purpose of the voltage bias is to excite the system away from its thermo-chemical equilibrium. We emphasize that the perturbation could be different in practice and still similar intrinsic dynamics would show up, irrespective of the specifics of the perturbation. It would also be feasible to use, e.g., short laser pulses which can perturb the system and then probe the consequent dynamics~\cite{Schmitt2008,Matsunaga2013,Kemper2015,Sentef2017,Werdehausen2018}. Nonetheless, the theoretical description of these processes is a challenge as these nanoscale devices are operating at high frequencies (THz) so the systems do not necessarily relax to a steady-state configuration instantly. In contrast, there are transient effects depending on, e.g., the system's geometry or topological character~\cite{Khosravi2009,Perfetto2010,Vieira2009,Wang2015,daRocha2015,Tuovinen2019}, its predisposition to external perturbations or thermal gradients~\cite{Kurth2010,Arrachea2010,Arrachea2012,Ness2011,Tuovinen2016PRB,Covito2018}, and the physical properties of the transported quanta and their mutual interactions~\cite{Wijewardane2005,Myohanen2008,Uimonen2011,Myohanen2012ic,Latini2014,Verdozzi2006,Tuovinen2018}. There is an increasing demand for theoretical and computational tools capable of addressing, in a general but computationally tractable level, the time-dependent responses of nanomaterials.

In this paper, we investigate how impurity invisibility in graphene~\cite{Duffy2016,Ruiz2016} takes place in the transient regime. Depending on the dopant conformation with respect to the underlying graphene nanoribbon lattice symmetry, we identify whether the charge transport properties of the conducting device are modified due to impurity scattering. We observe that the time-resolved signals are highly sensitive to the impurity configurations: in the transient regime the charge-current pathways are reorganized due to the scattering introduced by the impurities. We argue that recognizing this mechanism is pivotal for efficient design of graphene-based ultrafast chemical sensing devices.

\section{Model and method}
\begin{figure}[t]
\centering
\includegraphics[width=0.475\textwidth]{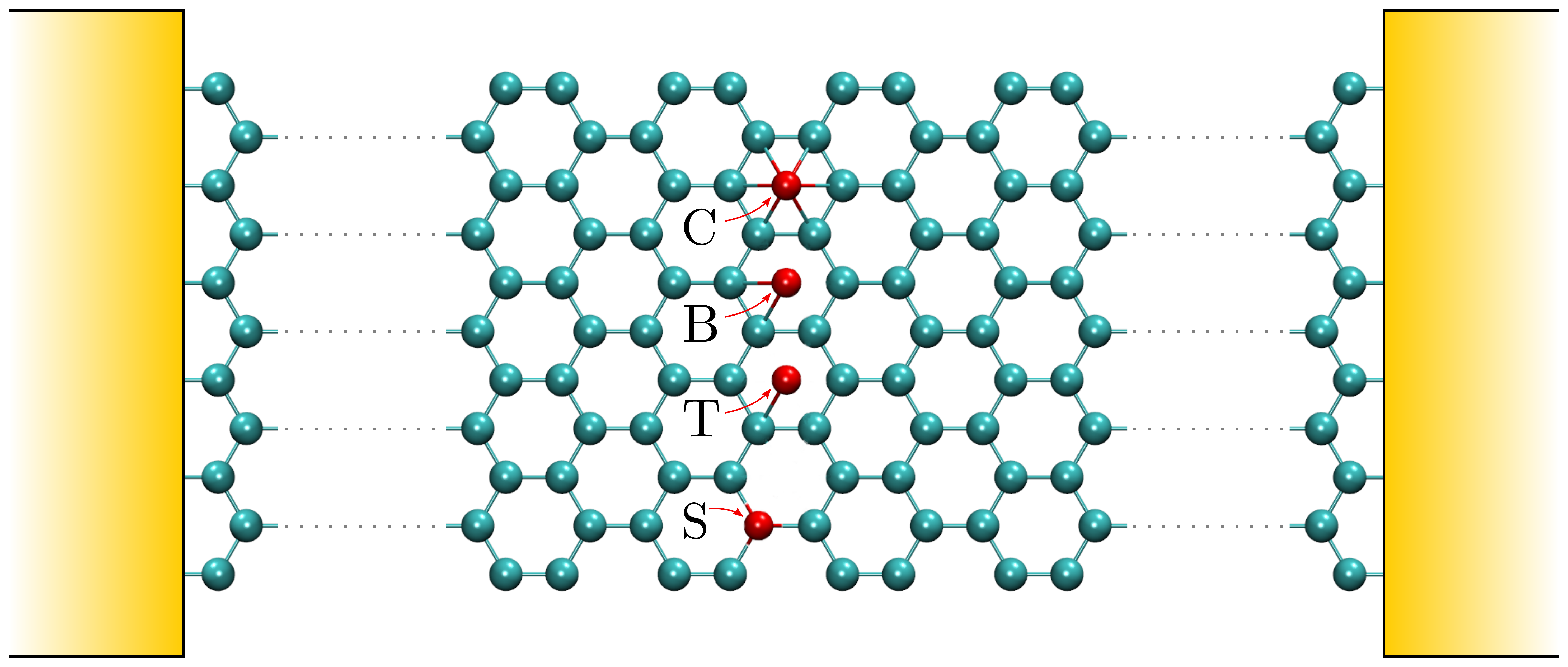}
\caption{Distinct impurity configurations on a GNR host (cyan atoms). Impurities (red atoms) can be positioned (i) right on the top of a carbon atom (T), (ii) positioned over a carbon-carbon bond characterizing a `bridge' configuration (B), (iii) placed on the center of an hexagonal ring (C), or (iv) substitute a carbon atom (S).}
\label{fig:zoom}
\end{figure}

We consider a system composed of metallic leads $\a$ connected to a central molecular structure, where we investigate the (charge) transport of noninteracting electrons. Even though, electron-electron and electron-phonon interaction should, in principle, influence the transport mechanisms, here we expect our noninteracting picture to be sufficient as recent studies on small monolayer graphene devices have revealed ballistic transfer lengths ranging from hundreds of nanometers to even micrometers at low temperatures~\cite{Miao2007,Lin2009}. The transport setup (cf. Fig.~\ref{fig:setup}) is \emph{partition-free}~\cite{Cini1980,Stefanucci2004,Ridley2018} which means that the whole system is initially contacted in a global thermo-chemical equilibrium at unique chemical potential $\mu$ and at temperature $(k_{\text{B}}T)^{-1} = \beta$. The central molecular structure is modeled by a tight-binding Hamiltonian
\be
\hat{H}_{\text{mol}} = \sum_{mn} T_{mn}\hat{c}_m^\dagger \hat{c}_n ,
\ee
where $T_{mn}$ accounts for hoppings between the lattice sites $m$ and $n$. The operator $\hat{c}_n$ ($\hat{c}_m^\dagger$) annihilates (creates) an electron on site $n$ ($m$) of the host lattice. In practice, we consider GNRs as the central molecular structure, and we set $T_{mn}=-\gamma$ for nearest neighbours $m$ and $n$, and $\gamma=2{.}7$~eV. Second and third nearest neighbour hoppings could be included similarly~\cite{CastroNeto2009,Harju2010} but here we consider particle-hole symmetric cases and only take the first nearest neighbours into account. For impurities in the central conducting device we have similarly
\be\label{eq:imp}
\hat{H}_{\text{imp}} = \sum_n [\eps_{\text{imp}} \hat{c}_n^\dagger \hat{c}_n + \g_{\text{imp}} (\hat{c}_{a_n}^\dagger \hat{c}_n + \hat{c}_n^\dagger \hat{c}_{a_n})],
\ee
where the index $n$ runs over the impurities, and $a_n$ is the host site in the (pristine) central region where the impurity is attached to. The operator $\hat{c}_{j}$ ($\hat{c}_{j}^\dagger$) annihilates (creates) an electron on site $j$ that can be on the host lattice or impurity site. Parameters for the on-site  ($\eps_{\text{imp}}$) and hopping energies ($\g_{\text{imp}}$) for the impurities can be related to ab initio calculations~\cite{Robinson2008,Wehling2010} following Density Functional Theory (DFT). The impurity configurations considered here are shown in Fig.~\ref{fig:zoom}. An impurity can be positioned (i) right on the top of a carbon atom (T), (ii) positioned over a carbon-carbon bond characterizing a ``bridge'' configuration (B), (iii) placed on the center of an hexagonal ring (C), or (iv) substitute a carbon atom, yielding a ``substitutional'' configuration (S). The leads are described as (semi-infinite) reservoirs
\be\label{eq:hlead}
\hat{H}_{\text{lead}} = \sum_{k\a} \eps_{k\a} \hat{c}_{k\a}^\dagger \hat{c}_{k\a}
\ee
with $\eps_{k\a}$ corresponding to the energy dispersion for basis states $k$ in lead $\a$. For a one-dimensional tight-binding structure this is given by $\eps_{k\a} = 2t_\a \cos k$ with $t_\a$ the hopping energy between the lead's lattice sites. Here we assume that the density of states of the leads is smooth and wide enough allowing us to consider the wide-band approximation (WBA). In WBA the lead density of states is assumed independent of energy, which in practice means that we choose the energy scales in the leads much higher than other energy scales in the central region. As we concentrate on the effects between the graphene nanoribbon and the impurities this allows us to neglect the precise description of the electronic structure of the leads. This is further justified in typical transport setups where the bandwidth of the leads is sufficiently large (e.g. gold electrodes) compared to the applied bias voltage~\cite{Zhu2005,Verzijl2013,Covito2018,Ridley2019}. The leads are connected to the central region by the coupling Hamiltonian
\be
\hat{H}_{\text{coupl}} = \sum_{m,k\a} (T_{m,k\a} \hat{c}_m^\dagger \hat{c}_{k\a} + \hc ).
\ee
with $T_{m,k\a}$ the hopping energy coupling lead's states with the lattice sites of the molecular structure. The total Hamiltonian is then combined as $\hat{H}_{\text{tot}} = \hat{H}_{\text{mol}} + \hat{H}_{\text{imp}} + \hat{H}_{\text{lead}} + \hat{H}_{\text{coupl}}$. We consider a switch-on of a bias voltage $V_\a$ in lead $\a$ at time $t=0$ meaning that the lead energy dispersion in Eq.~\eqref{eq:hlead} becomes $\eps_{k\alpha}\to \eps_{k\a} + V_\a$. Due to this nonequilibrium condition, charge carriers start to flow through the molecular conducting channel, in our case, a GNR. We stress that the coupling matrix elements between the central region and the leads, $T_{m,k\a}$, are constant at all times as the system is partition-free. Here we consider only voltage biases as mean of perturbation but recently it has also been shown that temperature gradients may be included in this consideration at an equal footing~\cite{Eich2014,Eich2016,Covito2018}.

In the literature, a considerable amount of works uses the method of Landauer and B{\"u}ttiker~(LB)~\cite{Landauer1970,Buttiker1986} to determine the transport properties of nanoscale devices as it provides a very simple and intuitive physical picture of the transport mechanism. The current $I_{\alpha\delta}$ in lead $\delta$ is calculated from the scattering states originating from lead $\alpha\neq\delta$. These scattering amplitudes are typically written as transmission probabilities for an electron to traverse from lead $\alpha$ to lead $\delta$. The stationary current in lead $\delta$ is obtained from the difference $\sum_{\alpha\neq\delta} \lbrack I_{\alpha\delta}-I_{\delta\alpha} \rbrack$. Here we also address time-resolved currents, and the nonequilibrium Green's function (NEGF) formalism~\cite{svlbook} provides a natural framework for this as it is not limited to the stationary state. We describe the time-evolution (transient information) of the system by the NEGF formalism where the Hamiltonian, the Green's function, and correlation effects (self-energies) are coupled by the integro-differential Kadanoff-Baym equations of motion~\cite{svlbook}. For the model system described above, an analytic solution for the time-dependent one-particle density matrix of the central region and for the time-dependent current between the central region and the leads can be found~\cite{Tuovinen2014,Tuovinen2016}. This time-dependent extension to the LB formalism (TD-LB) shares the simple interpretation of the original LB formalism and does not increase the computational cost as it would be the case if one solves numerically the complete Kadanoff-Baym set of equations~\cite{Dahlen2007,Stan2009}. In addition, an arbitrary time-dependence may also be included in the bias voltage, e.g., ac driving~\cite{Ridley2015,Ridley2017}. We emphasize that our method allows for studying the transient and stationary regimes at an equal footing since the stationary LB formula is recovered at the long-time limit $t\to\infty$.

\section{Results}\label{sec:results}

%\subsection{Graphene nanoribbon samples}

We consider GNRs of varying widths with armchair edges in the transport direction (see Figs.~\ref{fig:setup} and~\ref{fig:zoom}). The widths of the GNRs studied here are $N=\{11,12\}$ with $N$ indicating the number of carbon-dimers across the ribbon width representing, respectively, the metallic and semiconducting families of armchair GNRs: $N=3p-1$ and $N=3p$ with $p$ an integer number~\cite{Prezzi2008,Kimouche2015}. In addition to the pristine GNR, we consider impurities being adsorbed or substitutionally placed over the GNR host as shown in Fig.~\ref{fig:zoom}. The impurities can connect to the pristine GNR in four different configurations: `Center' (C), `Bridge' (B), `Top' (T), and `Substitutional' (S)~\cite{Duffy2016}. For the impurities we set $\eps_{\text{imp}}=0.66\gamma$ and $\gamma_{\text{imp}}=-2.2\gamma$ in Eq.~\eqref{eq:imp}~\cite{Robinson2008}. The left-most atoms of the graphene nanoribbon are connected to the left lead and the right-most atoms to the right lead, see Fig.~\ref{fig:zoom}. The length of the GNR in the transport direction is $10$ hexagons ($\approx 4$~nm) which is large compared to the impurity section. This choice is justified also because the overall transient features have been shown to scale with the length of the GNRs~\cite{Tuovinen2014}. For the sake of simplicity, we study transient responses when the GNR is subjected to adsorption of $4$ impurities; these correspond to the four centermost hexagons in the GNR (Fig.~\ref{fig:zoom}). More complex chemical perturbations such as increase in the number of impurities, asymmetric bonds, random distribution of impurities etc.~\cite{Fuerst2009,Ervasti2015,Brun2016} could also be addressed with the same theoretical toolbox. However, our understanding can benefit from such simplified picture that enables us to address this problem with mathematical transparency. In this way, we can identify with clarity on typical transport patterns and universal responses in a class of impurities that attach to the graphene lattice following these particular bonding symmetries.

%\subsection{Current--voltage characteristics}

\begin{figure}[t]
\centering
%\showthe\columnwidth
\includegraphics[width=0.465\textwidth]{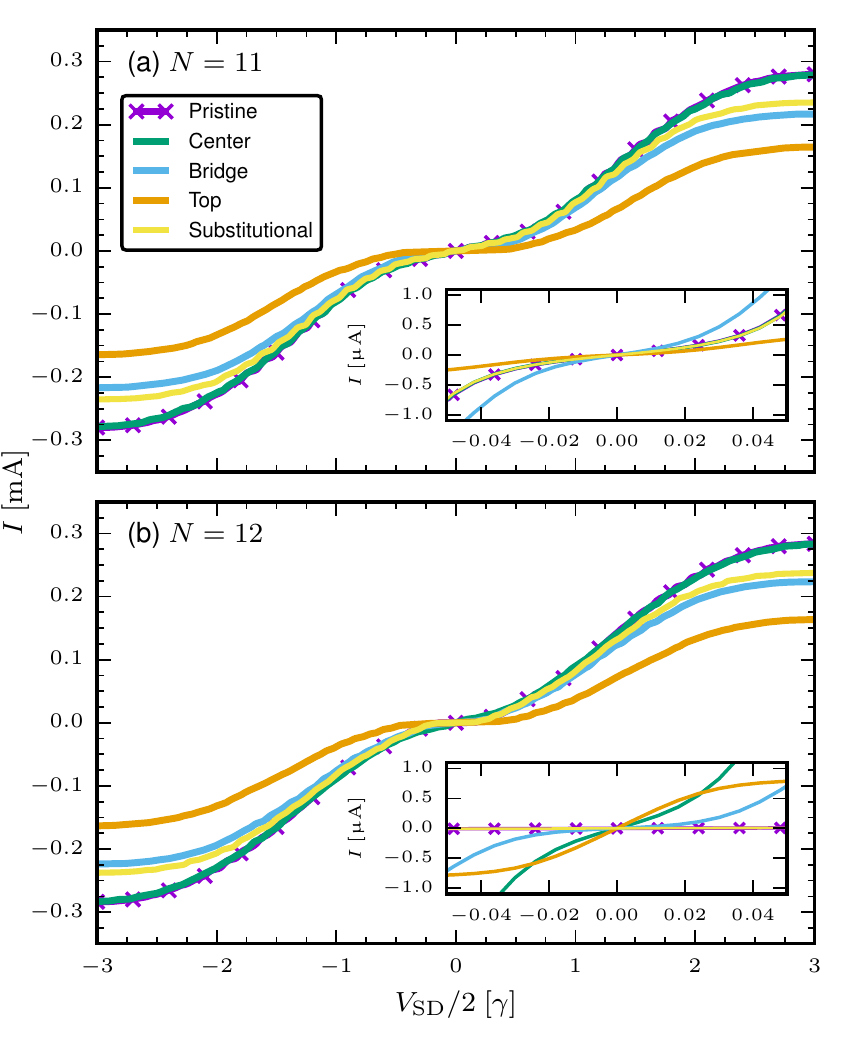}
\caption{Current--voltage characteristics of a (a) $N=11$ and (b) $N=12$ GNR in pristine and adsorbed/doped forms. Adatoms are placed on the center, bridge, top, and substitutional configurations as specified in Fig.~\ref{fig:zoom}. The insets show a zoom-in at the low-voltage regime.}
\label{fig:ivcurves}
\end{figure}

\emph{Current--voltage characteristics.---} We start by setting a source-drain voltage, $V_{\text{SD}}\equiv V_L - V_R$, over a GNR section sandwiched by left (L) and right (R) leads and evaluate the stationary current with $I=(I_L+I_R)/2$ where $I_L$ and $I_R$ are the currents through the left and right interface, respectively. This gives $I$-$V$ curves as seen in Fig.~\ref{fig:ivcurves}. From the current--voltage characteristics, we see that placing the impurities on the center configuration, in general, corresponds to impurity-invisibility~\cite{Duffy2016} as the curves are essentially on top of the pristine ones. We have tested with different on-site and hopping energy parameters for the impurity that this symmetric configuration gives rise to vanishingly small scattering regardless of the specifics of the impurity. The only deviation between the pristine and center configurations is seen in a very small bias voltage window in the $N=12$ ribbon (see inset in Fig.~\ref{fig:ivcurves}(b)). This effect could be related to Anderson localization, due to disorder induced by the impurities, leading to a metal-insulator transition~\cite{Garcia2014,Irmer2018}. In general, from Fig.~\ref{fig:ivcurves} we confirm that the P $N=11$ ribbon is metallic (nonzero slope at zero bias), and doping with the B configuration makes the ribbon more metallic-like while doping with the T configuration makes the ribbon more semiconducting-like. Also, the P $N=12$ ribbon is semiconducting and the other configurations show that the absolute values of the stationary currents are smaller, i.e., the impurities introduce scattering to some extent.

%\subsection{Current transients}

\begin{figure}[t]
\centering
\includegraphics[width=0.465\textwidth]{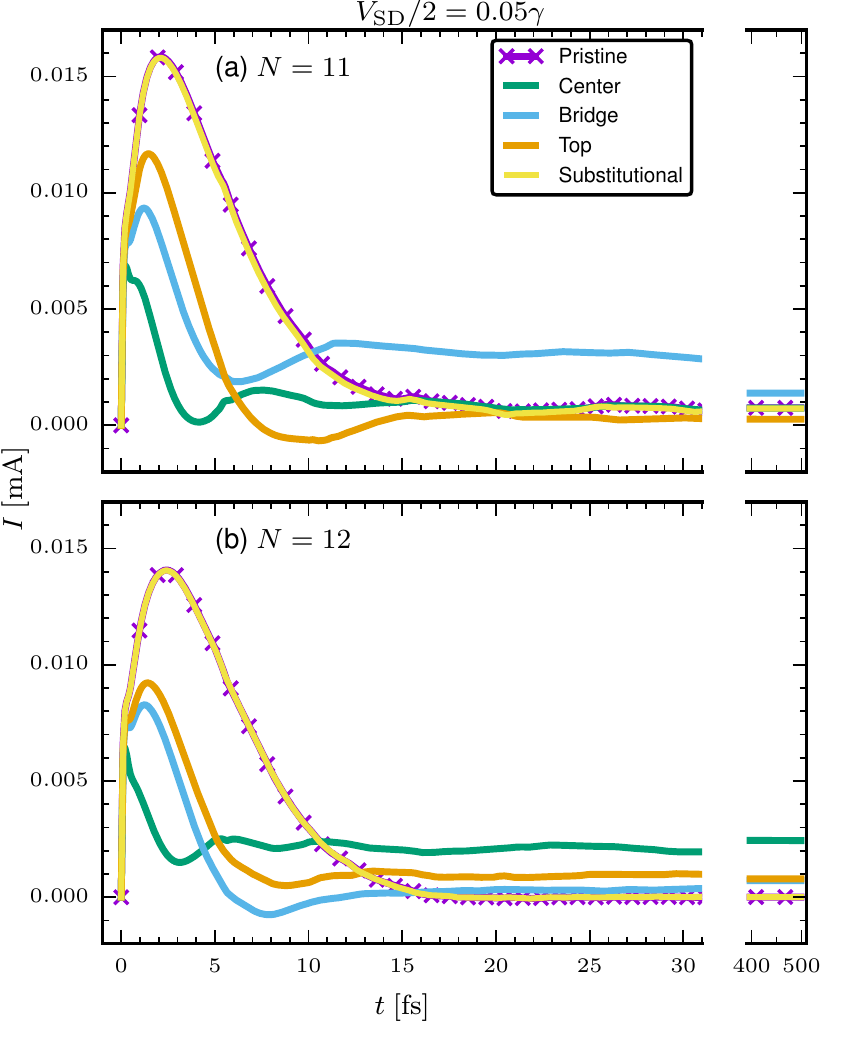}
\caption{Time-dependent currents driven by a small bias voltage through the (a) $N=11$ ribbons and (b) $N=12$ ribbons. The long-time limit of the currents is shown as a cutout on the right-hand side.}
\label{fig:tdcurrents}
\end{figure}

\begin{figure}[t]
\centering
\includegraphics[width=0.465\textwidth]{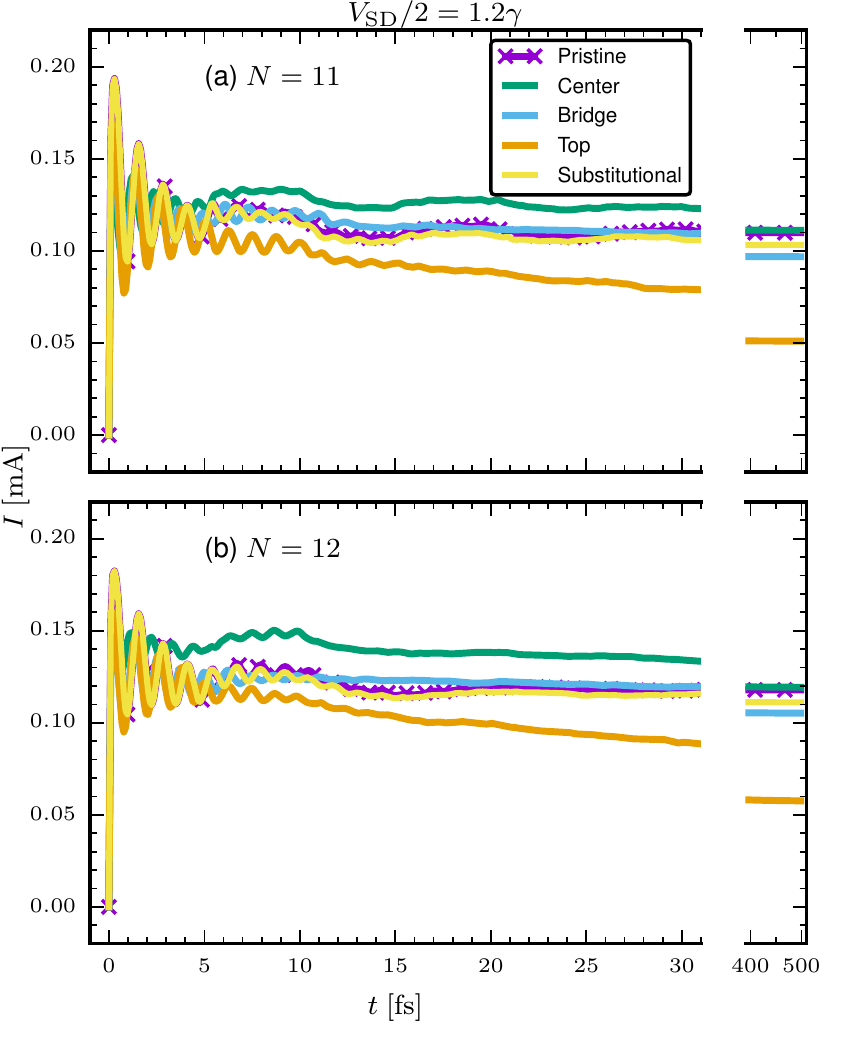}
\caption{Same as Fig.~\ref{fig:tdcurrents} but with a large bias voltage.}
\label{fig:tdcurrents2}
\end{figure}

\emph{Current transients.---}Now we investigate, how the stationary state in Fig.~\ref{fig:ivcurves} is reached from the transient regime. As we observed some albeit small different behavior at small and large voltages, we performed transient calculations also in these two regimes by fixing the bias voltage to $V_{\text{SD}}/2=0.05\g$ and $V_{\text{SD}}/2=1.2\g$, respectively. The time-dependent current signals are shown in Fig.~\ref{fig:tdcurrents} where we depict the initial transient behaviour (up to $30$~fs) and also the long-time limit (up to $500$~fs) at which we observed saturation of the currents. At small voltages there is considerable difference between the P and C configurations during the transient. As expected from Fig.~\ref{fig:ivcurves}, they saturate to the same value for the $N=11$ ribbon and to a different value for the $N=12$ ribbon. Larger voltages (Fig.~\ref{fig:tdcurrents2}) bring more transient features but we see that in both $N=11$ and $N=12$ ribbons the P and C configurations saturate essentially to the same value, although during the transient they can be very different. At higher voltages some configurations (T and B) take very long times to saturate (hundreds of femtoseconds); these impurity states introduce a considerable amount of back-and-forth scattering, which is not coupled to the leads, and these states have a long lifetime resulting in very slow damping.

%\subsection{Local charge densities and bond currents}

\begin{figure*}[ht!]
\centering
\includegraphics[width=\textwidth]{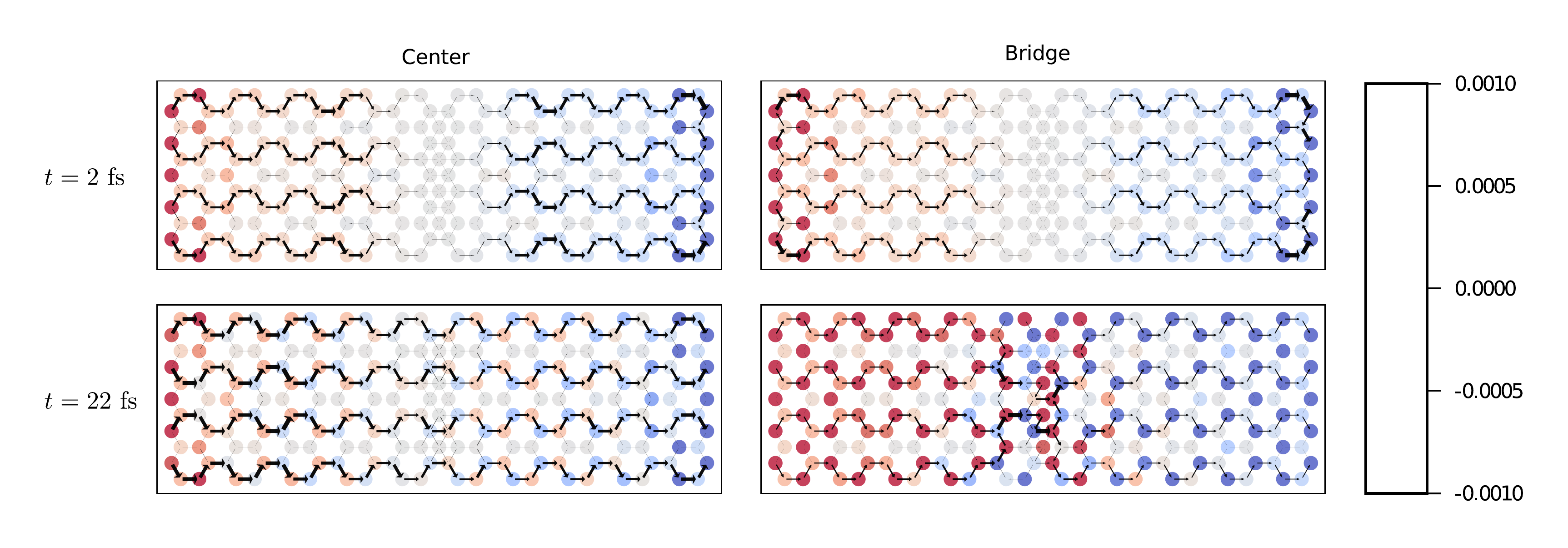}
\caption{Snapshots of the local charge densities and bond currents along the nanoribbons during the initial transient due to a small bias voltage, $V_{\text{SD}}/2=0.05\g$. The charge densities are calculated as the difference from the ground-state density (color map). The strength of the bond current is indicated by the width of the black arrows.}
\label{fig:snapshot}
\end{figure*}

\begin{figure*}[ht!]
\centering
\includegraphics[width=\textwidth]{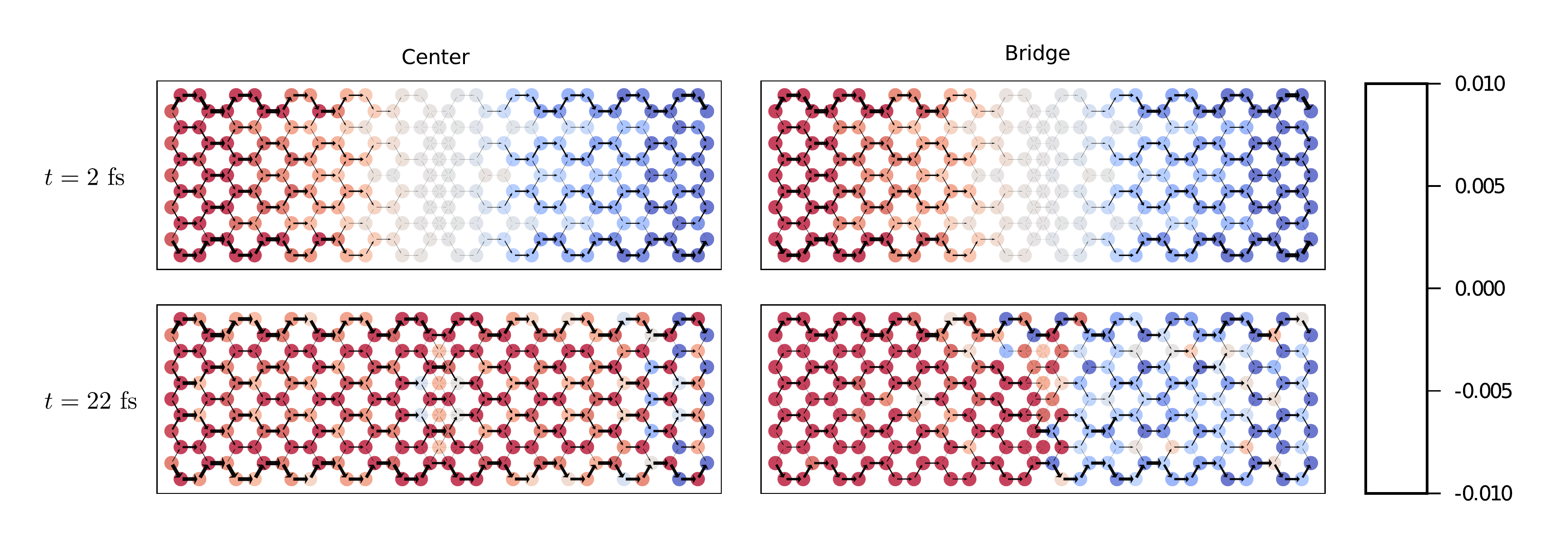}
\caption{Same as Fig.~\ref{fig:snapshot} but with a large bias voltage, $V_{\text{SD}}/2=1.2\g$. (Note the different scale on the color bar.)}
\label{fig:snapshot2}
\end{figure*}

\emph{Local charge densities and bond currents.---}As we have seen above, even if the time-dependent currents through the nanoribbons saturate to expected values from stationary calculations, during the transient the currents oscillate significantly. In addition to the interface currents between the nanoribbons and the leads, we now investigate the local charge fluctuations and bond current patterns within the entire samples. In order to access this information, we need to evaluate the full one-particle density matrix, where the diagonal elements correspond to the local site densities and the off-diagonal elements correspond to the bond currents between the sites~\cite{Tuovinen2014}. We concentrate on the initial transient to understand how the role of impurities affects the formation of the stationary state, and we focus our discussion on two representative impurity configurations, C and B. Complete results are shown in the Supplementary Information~\cite{SM}.

From the time-dependent currents in Figs.~\ref{fig:tdcurrents} and~\ref{fig:tdcurrents2} and from the snapshots in Figs.~\ref{fig:snapshot} and~\ref{fig:snapshot2} we see that, even though the stationary current through the C configuration is mostly unaffected by the impurity sites, in the transient regime the impurities provide a ``shock absorber'' for the initial density wavefront. In the C configuration the density wavefronts undergo a symmetry-driven destructive interference, and the opposing bond currents cancel each other. This transparency is lifted once the lattice symmetry of the system is broken in other impurity configurations. This effect is observed by the decreased initial current peak at small voltages in Fig.~\ref{fig:tdcurrents}, and as modified transient oscillations at high voltages in Fig.~\ref{fig:tdcurrents2}. The snapshots in Figs.~\ref{fig:snapshot} and~\ref{fig:snapshot2} show the density variation (with respect to the ground state) and bond-current profiles before the first collision of the density wavefronts at the middle of the ribbons ($t=2$~fs), and later when the wavepackets are reflecting from the lead interfaces back to the middle of the ribbons ($t=22$~fs). We see how the B configuration introduces a peculiarly locked current pattern around the impurity atoms. This effect results in a remarkably partitioned charge distribution compared to the C configuration.

The full dynamics is better visualized by animations in the Supplementary Information~\cite{SM}. From the animations we may also see how the overall current pathways through the ribbon are modified by the impurities in real time. We note in passing that we observe, similar to Ref.~\cite{Duffy2016}, that the top-bonded impurities are strong scatterers compared to other impurity configurations, and the local bond-current profiles significantly rearrange and focus due to the impurities~\cite{SM,daRocha2015}.

%\subsection{Charge pumping}\label{sec:pump}

\begin{figure}[t]
\centering
\includegraphics[width=0.465\textwidth]{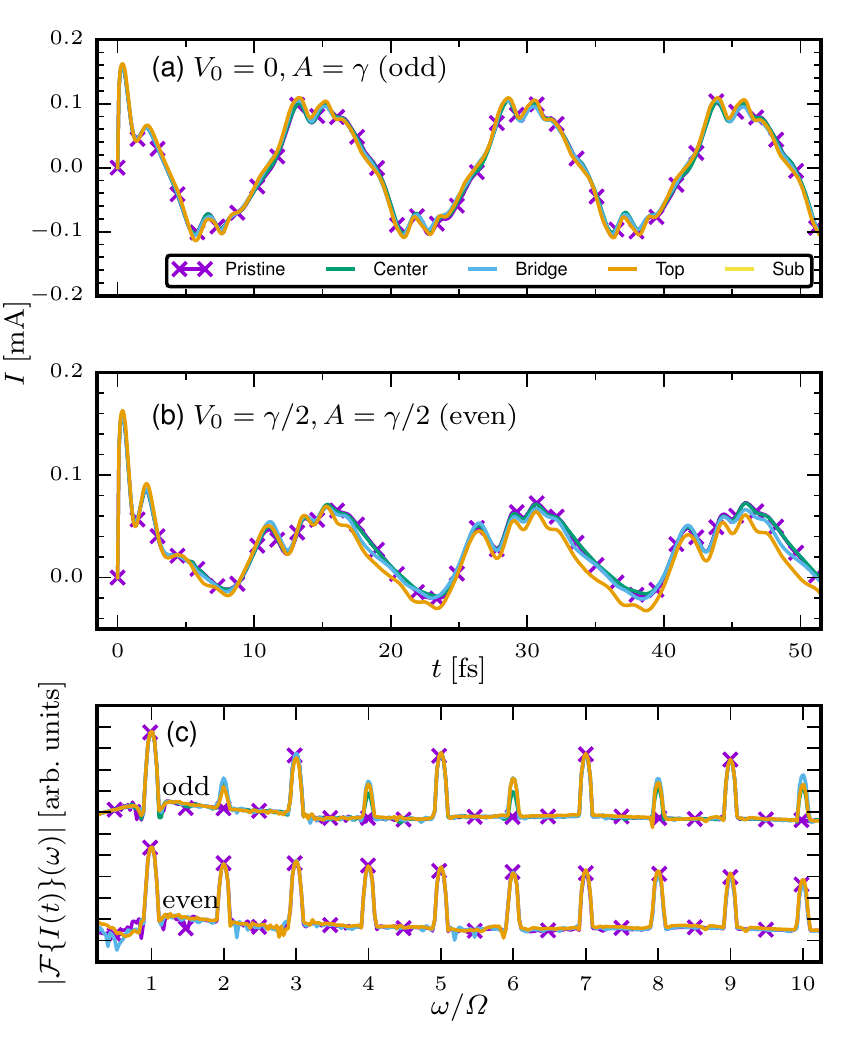}
\caption{Time-dependent currents through the $N=11$ ribbons driven by ac bias voltage. Current responses to (a) an odd-inversion-symmetric drive, and (b) a broken-inversion-symmetric drive. (c) The corresponding Fourier transforms.}
\label{fig:pumpcurrent}
\end{figure}

\emph{Transient charge pumping.---}We have seen above how breaking the lattice symmetry of the unbiased graphene sample by introducing impurities leads to different signals both in the transient and stationary regimes. The symmetry of the transport setup can also be broken by the driving mechanism, and thus, we consider charge pumping through the graphene samples~\cite{daRocha2010,FoaTorres2011,Connolly2013,Jnawali2013,Zhang2013,Ridley2017,Wang2018} in the transient regime. In contrast to the previous sections, we now introduce a harmonic driving $V_L = -V_R = V(t)$ with the voltage profile
\be
V(t) = V_0 + A\cos(\varOmega t),
\ee
where $V_0$ is the source-drain dc voltage and $A$ the amplitude of the ac driving. We set the driving frequency to be $\varOmega=\g/10$ which corresponds to a period of $2\pi/\varOmega\approx 15$~fs. We consider two types of ac driving: (1) $V_0=0$, $A=\gamma$, i.e., ac driving around zero dc voltage with odd inversion symmetry of the voltage profile, $V(t+\pi/\varOmega)=-V(t)$; and (2) $V_0=\gamma/2$, $A=\gamma/2$, i.e., breaking the odd-inversion symmetry of the applied bias with a constant shift term.

In Fig.~\ref{fig:pumpcurrent} we show, for the $N=11$ ribbons, the current responses to the two different drives described above, and also the corresponding Fourier spectra. For better frequency resolution the Fourier transform is calculated from an extended temporal window up to $500$~fs, and Blackman-window filtering is used. We see from Fig.~\ref{fig:pumpcurrent}(a) and Fig.~\ref{fig:pumpcurrent}(c) that even if the time-dependent signals are essentially on top of each other for all the impurity configurations (within this temporal window), the frequency content is still different. The pristine ribbon expectedly excludes the even harmonics, showing pronounced peaks at $\w=(2n + 1)\varOmega$ only, due to the odd-inversion-symmetric drive. However, introducing any impurities (even in the C configuration) breaks the corresponding symmetry of the time-independent Hamiltonian, and peaks at even multiples ($\w=2n\Omega$) of the driving frequency also appear. In Fig.~\ref{fig:pumpcurrent}(b) the odd-inversion symmetry of the drive is already broken, so all the ribbons including the P show pronounced peaks in the Fourier spectra also at the even multiples of the driving frequency. In addition, we see that in both cases peaks up to very high harmonic order are visible, indicating operation far beyond the linear-response regime.

\section{Conclusions}\label{sec:concl}
We presented a time-resolved characterization of impurity invisibility in graphene nanoribbons. Our transport setup of graphene nanoribbons supplemented with impurities was described by a single $\pi$ orbital tight-binding framework where the impurity atoms were modeled by modified tight-binding parameters compared to the pristine graphene nanoribbons. We accessed the transport properties both at stationary and transient regimes by the TD-LB formalism, allowing for a fast and accurate simulation based on the NEGF method~\cite{svlbook,Tuovinen2016thesis}.

Our stationary results showed that the center-bonded impurities in graphene are invisible to conduction electrons being unable to scatter them~\cite{Duffy2016,Ruiz2016,Irmer2018}. We then compared the time-dependent build-up of a steady-state current after a sudden quench of the bias voltage for different impurity configurations, and we discovered that the dynamics for different configurations look significantly different. Further, our spatio-temporal-resolved results showed that the impurities induce rearrangement and focusing of the current pathways along the graphene nanoribbons. In addition to the stationary picture, we further argue that graphene nanoribbons could serve as excellent probes or chemical sensors via ultrafast transport measurements~\cite{Prechtel2012,Hunter2015,Rashidi2016,Karnetzky2018,McIver2018}.

Driving the graphene samples with strong ac bias voltage was shown to lead in highly nonlinear behavior. The resulting high-harmonic responses were shown to contain selective even-odd signals implying a generation of distinct on-off signals from an analog source which could be further realized as ac-dc conversion or rectification. These findings further highlight the great potential sensing devices probing ultrafast modifications in the sample, the general design of efficient circuitries, and engineering of, e.g., plasmonic and optical nanoscale devices~\cite{Baer2004,Ryzhii2012,Gao2014,Marinica2015,Lin2015}.

\acknowledgments
R.T. and M.A.S. acknowledge funding by the DFG (Grant No. SE 2558/2-1) through the Emmy Noether program.

%\bibliographystyle{apsrev4-1}
%\bibliography{refs}


%%%%%%%%%%%%%%%%%%%%%%%%%%%%%%%%%%%%%%%%%%%%%%%%%%%%%%%%%%%%%%%%%%%%%%%%%%%%%%%%
%%%%%%%%%%%%%%%%%%%%%%%%%%%%%%%%%%%%%%%%%%%%%%%%%%%%%%%%%%%%%%%%%%%%%%%%%%%%%%%%

% End document
\end{document}